\documentclass{PoS}

\usepackage{amsmath}

\title{Jets in non-equilibrium quark-gluon plasma}

\ShortTitle{Jets in non-equilibrium quark-gluon plasma}

\author{\speaker{Sigtryggur Hauksson}\\
        Department of Physics, McGill University, 3600 rue University, Montreal, QC, Canada H3A 2T8\\
        E-mail: \email{sigtryggur.hauksson@mail.mcgill.ca}}

\author{Sangyong Jeon\\
        Department of Physics, McGill University, 3600 rue University, Montreal, QC, Canada H3A 2T8\\
        E-mail: \email{jeon@physics.mcgill.ca}}
        
\author{Charles Gale\\
        Department of Physics, McGill University, 3600 rue University, Montreal, QC, Canada H3A 2T8\\
        E-mail: \email{gale@physics.mcgill.ca}}

\abstract{Jets are a promising way to probe the non-equilibrium physics of quark-gluon plasma (QGP). We study how an out-of-equilibrium medium induces a jet particle to emit gluons. Evaluation of the emission rate is complicated by Weibel instabilities which lead to an exponential growth of chromomagnetic fields. Deriving a quantum field theoretical description of an unstable QGP medium, we show that the chromomagnetic fields deflect jet particles during the gluon emission.}

\FullConference{International Conference on Hard and Electromagnetic Probes of High-Energy Nuclear Collisions\\
		30 September - 5 October 2018\\
		Aix-Les-Bains, Savoie, France}

\begin{document}

\section{Introduction}

Heavy-ion collisions are a doorway to the non-equilibrium physics of QCD. Far from equilibrium, highly occupied gluon fields in the initial stages evolve into a nearly thermalized quark-gluon plasma (QGP) medium which can be described using relativistic hydrodynamics. Furthermore, heavy-ion collisions give access to transport coefficients of the QGP which are fundamental quantities of QCD. These coefficients, which include shear and bulk viscosity, describe the medium's response to an external perturbation. A central goal is to learn about thermalization and the transport coefficients of QGP using a wide variety of experimental probes, preferably ones that are sensitive to the evolution of the medium, such as jets. This requires a detailed theoretical understanding of jet quenching in non-equilibrium QGP. 

\begin{figure}
    \centering
        \includegraphics[width=0.25\textwidth]{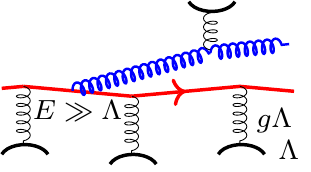}
    \caption{An energetic quark (red) emitting a gluon (blue) collinearly. During the emission both the quark and the gluon can interact arbitrarily often with medium through the exchange of soft gluons with energy \(g \Lambda\). Here \(\Lambda\) is the typical energy of particles in the medium and \(g\) is the coupling constant which we assume to be small.}
\label{Fig:LPM}
\end{figure}

At leading order in perturbation theory, an energetic jet particle interacts with a weakly coupled medium through two processes. Firstly, the jet particle can scatter off a quark or a gluon in the medium. Similar processes have been studied out-of-equilibrium in e.g. \cite{Schenke2006}. Secondly, the jet particle can emit a gluon collinearly when it gets transverse kicks from the medium making it slightly off shell, see Fig. \ref{Fig:LPM}. During the gluon emission the jet particle can get arbitrarily many such kicks. This repeated interaction with the medium acts coherently and tends to reduce the rate of emission, especially for more energetic jet particles. This is known as the Landau-Pomeranchuk-Migdal (LPM) effect \cite{AMYgluon}. In these proceedings we will study how collinear gluon emission can be evaluated in an out-of-equilibrium QGP. We will focus on challenges coming from the rapid growth of chromomagnetic fields in a non-equilibrium medium because of Weibel instabilities.

\section{Hard thermal loops and Weibel instabilities}

A weakly coupled plasma that is not too far from equilibrium \cite{AMYkinetic} has two types of excitations at leading order. Firstly, there are hard excitations of localized quasi-particles which can be described by kinetic theory using a momentum distribution \(f(\mathbf{k})\). Their momentum is of the order of \(\Lambda\) which corresponds to temperature in equilibrium. Secondly, there are soft excitations which are radiated by the hard particles and have momentum \(\sim g\Lambda\) where \(g\) is the coupling constant. They form a cloud of soft gluons which deflects the hard particles. The physics of the soft gluons is described by an effective theory called hard thermal loops (HTL). It was originally developed for a thermal medium \cite{Braaten1989} but has been extended to non-equilibrium media \cite{Mrowczynski2000}.


The retarded correlator in HTL,  \(G_{\mathrm{ret}}(\omega,\mathbf{q})\), characterizes how the cloud of soft gluons responds to fluctuations \cite{Romatschke2003}. In general, it has a pole in the upper half complex plane at \(\omega = i \gamma(\mathbf{q})\) because of Weibel instabilities \cite{Mrowczynski2016}.\footnote{These poles arise for any momentum distribution of hard particles, \(f(\mathbf{k})\), that is not isotropic. In general there are multiple instability poles but for simplicity we will assume that there is only one. Our results can easily be extended to the more general case.} The essence of these instabilities is that fluctuations produce strong colour currents which source exponentially growing gauge fields with growth rate \(\gamma\). Non-Abelian interaction of the gauge fields stops the growth at later times. Weibel instabilities have been proposed as being instrumental to the rapid thermalization in heavy-ion collisions \cite{Kurkela2011}. However, recent studies using classical field theory at very weak coupling suggest that the instabilities saturate before the system reaches a non-thermal fixed point making their evolution unimportant for later stages \cite{Berges2013}.

In non-equilibrium HTL effective theory, one seemingly gets a divergent rate for jet-medium interaction because of Weibel instabilities \cite{Baier2008}. In fact the same problem plagues photon production, heavy-quark diffusion, and the kinetic theory of quarks and gluons.\footnote{We note that calculations with a classical probe are well defined, see \cite{Carrington2016}} Thus it is vital to understand instabilities in HTL effective theory, irrespective of how big a role they play in the initial stages of heavy-ion collisions. In these proceedings we show that HTL is indeed a well-defined framework for non-equilibrium systems. This allows for calculations of jet-medium interaction which can be used to extract transport coefficients of the QGP using jet observables.

%

%

\section{Unstable plasmas}

The  unstable plasma in HTL effective theory cannot be considered as static. Instead we must follow the time evolution of the plasma starting at initial time \(\tau=0\). We need analytic two-point correlators representing how the cloud of soft gluons changes in time in order to evaluate quantum effects like the LPM effect. Firstly, there is the retarded correlator, \(G_{\mathrm{ret}}\), which has been derived in earlier works \cite{Romatschke2003}. Its poles describe dispersion relations of soft modes in the plasma. Secondly, we need the classical \(rr\) correlator which is defined as \(G_{rr}^{\mu\nu} = \frac{1}{2}\langle \{A^{\mu},A^{\nu}\}\rangle.\)  
It gives the occupation of modes in the plasma. In thermal equilibrium it is given by \(G_{rr}(P) = (\frac{1}{2} + f_B(p^0)) \left[ G_{\mathrm{ret}} - G_{\mathrm{adv}} \right]\) where \(G_{\mathrm{adv}}(P) = -G_{\mathrm{ret}}(P)^*\) is the advanced propagator and the Bose-Einstein distribution, \(f_{B}\), describes occupation in equilibrium. In an unstable plasma we expect the \(rr\) correlator to 
grow exponentially in time because of instabilities.

We specify initial conditions for the unstable plasma through the momentum distribution of hard particles, \(f(\mathbf{k})\).
The first instances in the evolution of the plasma are complicated because of correlations with the initial state. We thus wait until time \(\tau \gg 1/g^2 \Lambda\) where \(\Lambda\) is a hard scale determined by the momentum distribution.  We also assume for simplicity that there are no soft gluons at the initial time; they are built up by radiation from the hard particles. Furthermore, the growth of the gauge field ceases at late times because of its non-Abelian self-interaction. Then the strength of the field becomes \(\exp(\gamma \tau) \sim 1/g^2\) which goes beyond HTL effective theory. The growth rate is \(\gamma \sim \xi g \Lambda\) where the anisotropy, \(\xi\), measures how deformed \(f(\mathbf{k})\) is from an isotropic distribution. Thus, we must consider times \( (g^2 \Lambda)^{-1} \ll \tau \ll (\xi g \Lambda)^{-1}\). Clearly, we have to assume that \(\xi \ll g\); in other words we need a small anisotropy if the time evolution is to be slow enough for a controlled calculation. Despite the small anisotropy, the chromomagnetic fields will grow to large values. 


Using the above approximations\footnote{We furthermore only include leading order effects in \(\xi\) and ignore terms which oscillate rapidly during the time it takes to emit a gluon, \(1/g^2 \Lambda\), since the oscillations cancel out.} one gets analytic expressions for the correlators \cite{inprep}. We write the retarded correlator as 
\begin{equation}
G_{\mathrm{ret}}(Q) = \widetilde{G}_{\mathrm{ret}}(Q) + \frac{D(\mathbf{q})}{q^0 - i\gamma(\mathbf{q})}
\end{equation} 
where we have isolated the instability part and \(\widetilde{G}_{\mathrm{ret}}\) has no poles in the upper half complex plane.
Then the \(rr\) propagator can be written as \(G_{rr} = \widetilde{G}_{rr} + G_{rr}^{\mathrm{inst}}\). The first term, \( \widetilde{G}_{rr} = \widetilde{G}_{\mathrm{ret}} \; \Pi_{aa} \; \widetilde{G}_{\mathrm{adv}}\), gives the occupation density of the fluctuating soft gluon cloud which is present in thermal equilibrium and does not evolve in time.\footnote{Here \(\widetilde{G}_{\mathrm{adv}} = \widetilde{G}_{\mathrm{ret}}^{*}\) comes from the advanced propagator and \(\Pi_{aa}\) is one component of the polarization tensor.} The second term can be written as
\begin{equation}
G_{rr}^{\mathrm{inst}}(x^0,y^0;\mathbf{q}) = \left( D\; \Pi_{aa} \; D^{\dagger} \right) \frac{1}{2\gamma} \left[e^{2\gamma \tau} - e^{-\gamma |t|} \right]
\end{equation}
in the time domain with three-momentum. It describes how the gauge field grows exponentially in time \(\tau = \frac{x^0 + y^0}{2}\) because of the instabilities. Here \(t = x^0 - y^0\) is the variable conjugate to frequency.\footnote{This can be seen from the Fourier transform of the propagator which is defined as \\
\(
G_{rr}^{\mathrm{inst}}(\omega,\mathbf{q};\tau) = \int d(x^0 - y^0) \; e^{i\omega(x^0-y^0)} G_{rr}^{\mathrm{inst}}(x^0,y^0;\mathbf{q}).
\)}
 We can now see why earlier calculations got a divergent rate for jet-medium interaction: Ignoring evolution in time, they did not include the term \(e^{2\gamma \tau}\). This leads to a divergence for modes that grow slowly, \(\gamma \rightarrow 0\). Here that limit is perfectly well-defined and the occupation density of soft gluons is always finite.  
 
\section{Jet propagation in an unstable plasma}

\vspace{-0.4cm}
\begin{figure}
    \centering
        \includegraphics[width=0.91\textwidth]{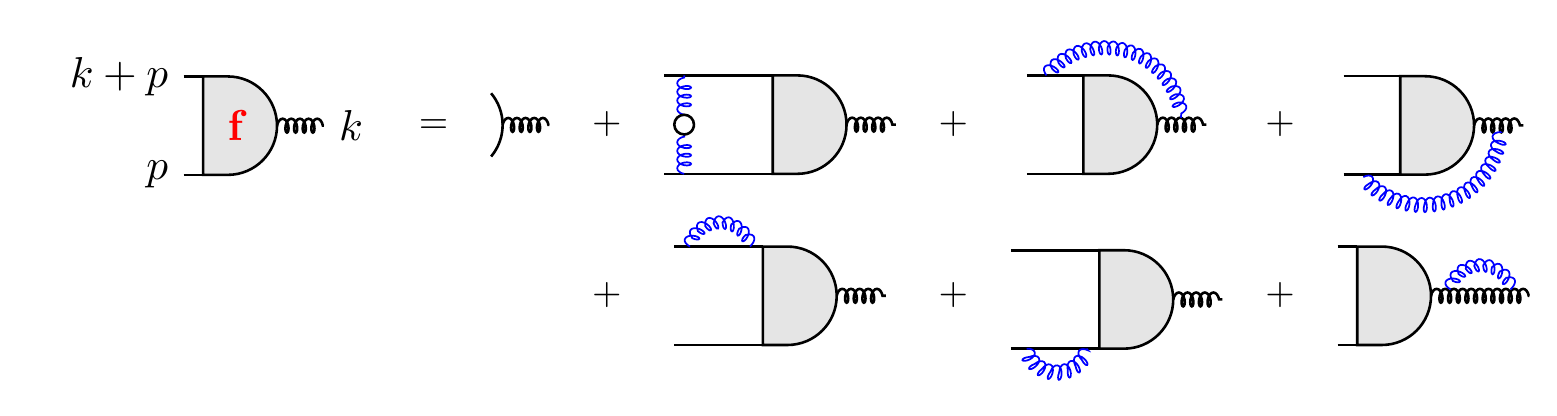}
    \caption{Diagrams needed to derive Eq. \eqref{Eq:int_eq}. They describe a quark with momentum \(k+p\) emitting a gluon with momentum \(k\) collinearly. The function \(\mathbf{f}\) denotes interaction with the unstable medium. The blue \(rr\) propagator in the diagrams includes contributions from the fluctuating soft gluon cloud through \(\widetilde{G}_{rr}\), as well as from the exponentially growing gluon fields, \(G_{rr}^{\mathrm{inst}}\).}
\label{Fig:int_eq}
\end{figure}
\vspace{-0.2cm}

Now that we have a description of an unstable plasma we can study how a non-equilibrium medium induces splitting of a jet particle. The particle gets kicks from the fluctuating soft gluon cloud as in thermal equilibrium. However, it is also deflected in the classical instability field which affects the emission rate. To study this further we need to evalute the Feynman diagrams shown in Fig. \ref{Fig:int_eq}. The function \(\mathbf{f}\) describes interaction with the medium and determines the rate, \(R\), of emitting gluons with momentum \(\mathbf{k}\) through
\begin{equation}
\frac{dR}{d^3\mathbf{k}} \sim \int d^2\mathbf{h} \; \mathbf{h} \cdot \mathrm{Re}\;\mathbf{f}(\mathbf{h}).
\end{equation} 
Here \(\mathbf{h}\) can be thought of as the transverse momentum particles acquire through interaction with the medium. (We are omitting momentum distributions and splitting functions in this equation, see \cite{AMYgluon} for further details.) 

The equation determining \(\mathbf{f}\) is \cite{inprep}
\begin{align} \label{Eq:int_eq}
\begin{split}
2\mathbf{h} = i \delta E \; \mathbf{f}(\mathbf{h}) \; + \; &\int_{\mathbf{q}_{\perp}} \mathcal{C}(\mathbf{q}_{\perp}) \left[ \mathbf{f}(\mathbf{h}) - \mathbf{f}(\mathbf{h} - k\, \mathbf{q}_{\perp})\right] + \;...\; \\
+ \; &\frac{i k}{\delta E + 2 i \Gamma_{\mathrm{eq}}} \mathbf{F} \cdot \nabla \; \mathbf{f}(\mathbf{h}) + \;...\; + \mathcal{O}(\partial_{\tau} \mathbf{f}). 
\end{split}
\end{align}
where the additional terms have the same form as the terms shown explicitly.
Here \(\delta E = E_p + E_k - E_{p+k}\) where the energies include thermal masses and \(\Gamma_{\mathrm{eq}}\) is the equilibrium decay width. The integral in the first line describes kicks the particles get from the fluctuating soft gluon cloud. The function \(\mathcal{C}(\mathbf{q}_{\perp}) = \int_{q^0} \int_{q^z} \; \widetilde{G}_{rr}(Q) \;2\pi \delta(q^0 - q^z)\) is the probability of getting a kick \(\mathbf{q}_{\perp}\) from the medium. It differs from the analogous function in thermal equilibrium. The second line describes deflection by the instability fields with
\begin{equation}
\mathbf{F} \sim g^2 \hat{K}_{\mu} \hat{K}_{\nu} \int d^3 q \left(D \;\Pi_{aa}\; D^{\dagger} \right) ^{\mu\nu} \gamma^{-1} \left[e^{2\gamma \tau} - 1 \right] \mathbf{q}_{\perp}
\end{equation}
where we have omitted some kinematic and color factors. This term grows exponentially in time with the instabilities  and deflects the particles continuously in one direction. There is furthermore a term \(\mathcal{O}(\partial_{\tau} \mathbf{f})\) which contributes at the same order. To go beyond leading logarithmic order in the coupling constant further diagrams need to be evaluated; this is the subject of future work. \\

 {\noindent }\textbf{Acknowledgment:} This work was supported in part by the Natural Sciences and Engineering Research Council of Canada. SH acknowledges support through grants from Fonds de Recherche du Qu\'ebec.

%

\bibliographystyle{elsarticle-num}
\bibliography{bibliography.bib}

\end{document}